\documentclass[journal=jacsat,manuscript=communication]{achemso}
\pdfoutput=1 
\usepackage[version=3]{mhchem} 
\usepackage{lineno,xcolor}

\usepackage{graphicx}
\usepackage[version=3]{mhchem} 
\usepackage{subfig}
\usepackage{dcolumn}
\newcolumntype{d}[1]{D{.}{\cdot}{#1} }
\usepackage{bm}

\usepackage{siunitx}

\title{Atomistic origins of high-performance in hybrid halide perovskite solar cells}

\author{Jarvist M. Frost}
\author{Keith T. Butler}
\author{Federico Brivio}
\author{Christopher H. Hendon}
\affiliation{Centre for Sustainable Chemical Technologies and Department of Chemistry, University of Bath, Claverton Down, Bath BA2 7AY, UK}

\author{Mark van Schilfgaarde}
\affiliation{Department of Physics, Kings College London, London WC2R 2LS, UK}

\author{Aron Walsh}
\email{a.walsh@bath.ac.uk}
\affiliation{Centre for Sustainable Chemical Technologies and Department of Chemistry, University of Bath, Claverton Down, Bath BA2 7AY, UK}


\begin{document}

\begin{abstract}

The performance of organometallic perovskite solar cells has rapidly surpassed that of both conventional dye-sensitised and organic photovoltaics. 
High power conversion efficiency can be realised in both mesoporous and thin-film device architectures.
We address the origin of this success in the context of the materials chemistry and physics of the bulk perovskite as described by electronic structure calculations. 
In addition to the basic optoelectronic properties essential for an efficient photovoltaic device (spectrally suitable band gap, high optical absorption, low carrier effective masses), the materials are structurally and compositionally flexible. 
As we show, hybrid perovskites exhibit spontaneous electric polarisation; we also suggest ways in which this can be tuned through judicious choice of the organic cation. 
The presence of ferroelectric domains will result in internal junctions that may aid separation of photoexcited electron and hole pairs, and reduction of recombination through segregation of charge carriers. 
The combination of high dielectric constant and low effective mass promotes both Wannier-Mott exciton separation and effective ionisation of donor and acceptor defects. 
The photoferroic effect could be exploited in nanostructured films to generate a higher open circuit voltage and may contribute to the current-voltage hysteresis observed in perovskite solar cells. 

\end{abstract}



\begin{figure*}[ht!]
\begin{center}
\resizebox{15 cm}{!}{\includegraphics*{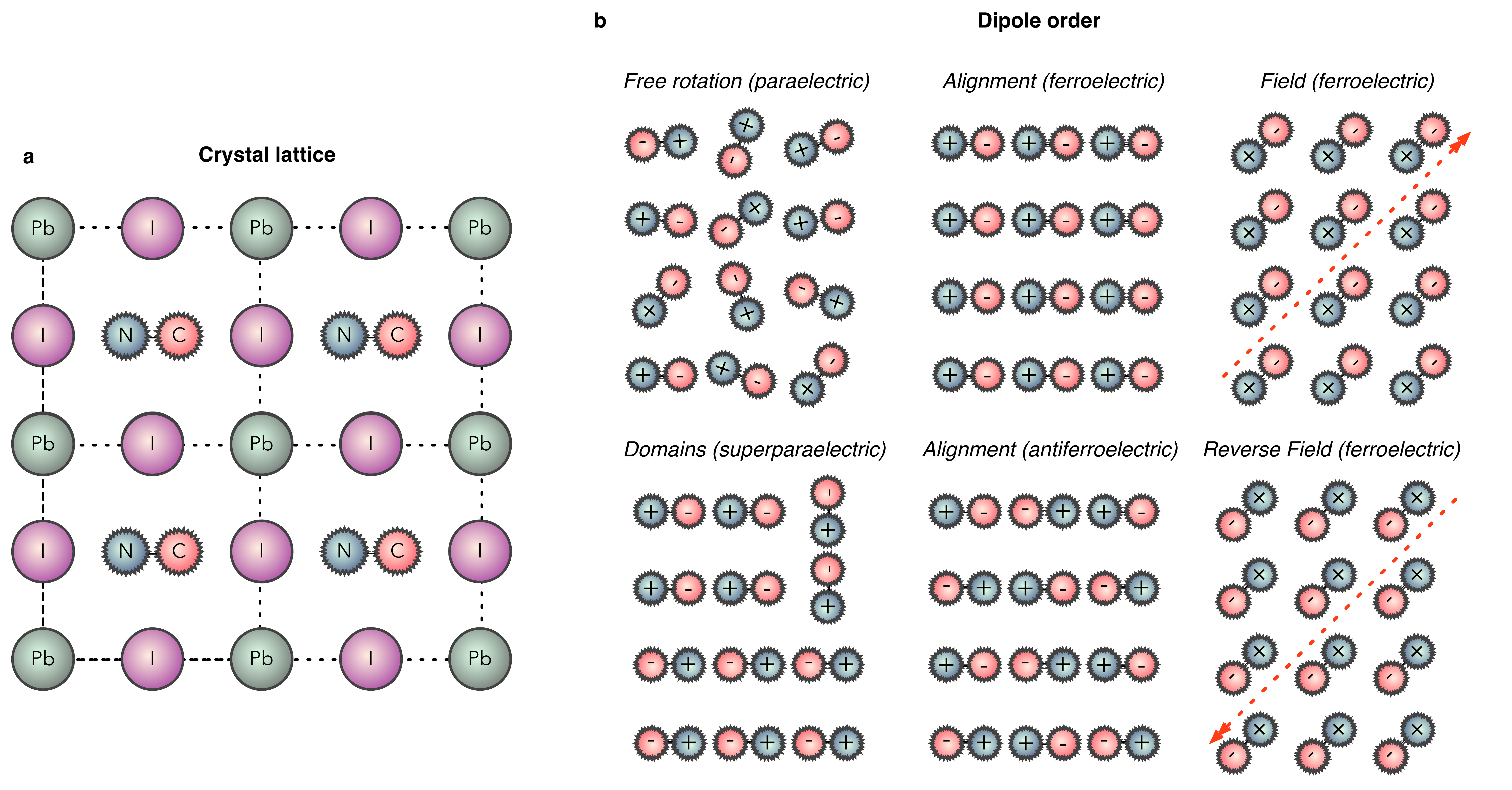}}
\caption{\label{fig-dip} Schematic perovskite crystal structure of MAPbI$_3$ (\textbf{a}), and the possible orientations of molecular dipoles within the lattice (\textbf{b}). Note that MA has an associated molecular dipole of 2.3 Debye, a fundamental difference compared to the spherical cation symmetry in inorganic perovskites such as CsSnI$_3$.} 
\end{center}
\end{figure*}

The development of halide pervoskite solar cells is as rapid as the technology is proving successful.\cite{kojima-6050,snaith-1,gratzel-1,gratzel-2,heo-486,snaith-2}
Research on this material system dates to the 1920s.\cite{wyckoff-349}
The most widely used hybrid perovskite solar cell material is methylammonium lead iodide (\ce{CH3NH3PbI3} or  \ce{MAPbI3}), where MA is a positively charged organic cation at the centre of a lead iodide cage structure (Figure \ref{fig-dip}).
Initially employed as a light sensitiser in mesoporous dye cells, these materials also function as an absorber and transport layer in a solid-state dye cell architecture, and most recently as the bulk material in a standard planar thin-film solar cell.\cite{park-2423}

Here we build upon recent computational studies
\cite{mosconi-13902,even-2999,even-2014,quarti-279,brivio-042111,brivio-2014,borriello-235214,yin-063903} 
and calculate the 
electrostatic cohesion energy; the electronic band energies; optical transitions; cation molecular polarisation tensors and energetic barriers for rotation; in hybrid lead iodide perovskites.
We show how lattice polarisation is affected by the polar organic cation, and how this gives rise to ferroelectric behaviour, which could enhance photovoltaic performance.  

\textbf{Perovskite Structure}
The crystal structure adopted by CaTiO$_3$ (the mineral \textit{perovskite}) is also found for a wide range of materials with ABX$_3$ stoichiometry, with two well studied cases being SrTiO$_3$ and BaTiO$_3$. Examples of insulating, semiconducting and superconducting perovskite structured materials are known; they represent a unique system class across solid-state chemistry and condensed matter physics. In particular, they are the archetypal systems for phases transitions with accessible cubic, tetragonal, orthorhombic, trigonal and monoclinic polymorphs depending on the tilting and rotation of the BX$_6$ polyhedra in the lattice.\cite{glazer-1972} Reversible phase changes can be induced by a range of stimuli including electric field, temperature and pressure.

There have been several reports of structural characterisation for single crystals of \ce{MAPbI3} and related compounds.\cite{stoumpos-9019,baikie-5628} Orthorhombic, tetragonal and cubic polymorphs have been identified, analogous to the inorganic perovskites. 
Indeed it has been suggested that for \ce{MAPbI3} an orthorhombic to tetragonal transition occurs at $\sim$ 160 K with a cubic phase being stable from around 330 K.\cite{baikie-5628} Disagreement between X-ray and electron diffraction has been noted, which suggests the presence of nanoscale structural domains.\cite{baikie-5628} 
This local disorder is also present in thin-film samples. The structure of thin-films within mesoporous titania has been investigated using X-ray scattering,\cite{choi-nl} which showed that the majority of the MAPbI$_3$ exists in a disordered state with a structural coherence length of only 14 \AA, i.e. the length of just two \ce{PbI3} cages.  
This behaviour will be explained below in the context of feedback between molecular and lattice polarisation---a ferroelectric effect.

\textbf{Perovskite Composition}
Within the formal stoichiometry of ABX$_3$, charge balancing can be achieved in a variety of ways. 
For metal oxide perovskites (ABO$_3$), the valence of the two metals must sum to six, i.e. I-V-O$_3$, II-IV-O$_3$ and III-III-O$_3$ perovskites are known with common examples being KTaO$_3$, SrTiO$_3$ and GdFeO$_3$. 
The range of accessible materials can be extended by partial substitution on the anion sub-lattice, e.g. in the formation of oxynitride and oxyhalide perovskites.\cite{castelli_computational_2012}  
With substitution on the metal sub-lattice, the so-called double, triple and quadruple perovskites are formed,\cite{gormezano-979} which may offer a route for property optimisation of hybrid compounds.

For halide perovskites, the valence of the two cations must sum to three, so the only viable ternary combination is I-II-X$_3$, e.g. CsSnI$_3$. In hybrid halide perovskites such as \ce{MAPbI3}, a divalent inorganic cation is present, with the monovalent metal replaced by an organic cation of equal charge. 
In principle, any singly charged molecular cation could be used, once there is sufficient space to fit it within the inorganic cage. 
If the size is too large, then the 3D perovskite structure is broken, as  demonstrated in the series of hybrid structures with lower dimensionality in the inorganic networks.\cite{calabrese-2328,mitzi-1,mitzi-2,borriello-235214} 
For layered structures, the crystal properties become highly anisotropic.  

The stability of ionic and heteropolar crystals, such as perovskites, is influenced by the Madelung electrostatic potential. 
The lattice energy and site electrostatic potentials is explored for each perovskite stoichiometry (Table \ref{tbl:properties}). These are calculated with an Ewald summation (in three dimensions) of the formal ion charges within the code \textit{GULP}\cite{gulp}. For group VI anions, the lattice energy decreases as the charge imbalance between the \textit{A} and \textit{B} sites is removed: a lower charge on the \textit{A} site is favoured. However, for group VII anions (i.e. halides) the electrostatic stabilisation is notably reduced, with a lattice energy of just -29.71 eV/cell and an electrostatic potential on the anion site ca. 50\% of the group VI anions. Due to this weaker, less-confining, potential, lower ionisation potentials (workfunctions) are expected for halide perovskites compared to, for example, metal oxides.\cite{scanlon-nmat,butler-2014}

\begin{table*}[ht]
\small
  \caption{Electrostatic lattice energy and site Madelung potentials for a range of ABX$_3$ perovskite structures (cubic lattice, \textit{a} = 6.00 \AA) assuming the formal oxidation state of each species. The hybrid halide perovskites are of type I--II--VII$_3$. Calculations are performed using the code GULP.}
  \label{tbl:properties}
  \begin{tabular}{ld{2}d{2}d{2}d{2}}
    \hline
    Stoichiometry & \multicolumn{1}{c}{$E_{lattice}$ (eV/cell)} & \multicolumn{1}{c}{$V_A$ (V)} & \multicolumn{1}{c}{$V_B$ (V)} & \multicolumn{1}{c}{$V_X$ (V)} \\
    \hline
I--V--VI$_3$ & -140.48 & -8.04 & -34.59 & 16.66 \\
II--IV--VI$_3$ & -118.82 & -12.93 & -29.71 & 15.49 \\
III--III--VI$_3$ & -106.92 & -17.81 & -24.82 & 14.33 \\
I--II--VII$_3$ & -29.71 & -6.46 & -14.85 & 7.75 \\
    \hline
\end{tabular}
\end{table*}

\textbf{From Inorganic to Hybrid Compounds}
An important distinction between inorganic and hybrid perovskites is the change from a spherically symmetric \textit{A} site (inorganic) to one of reduced symmetry (hybrid). The characteristic space groups of perovskite compounds are formally reduced. For example, the MA cation has the $C_{3v}$ point group and the associated highest-symmetry perovskite structure will be pseudo-cubic and not possess the inversion symmetry of its inorganic analogue. Thus uncertainty in assigned average diffraction patterns is not surprising. 

The presence of a polar molecule at the centre of the perovskite cage also introduces the possibility of orientational disorder and polarisation as drawn in Figure \ref{fig-dip}. 
A typical solid-state dielectric will exhibit a combination of fast electronic ($\epsilon_{\infty}$) and slow ionic ($\epsilon_{ionic}$) polarisation, which both contribute to the macroscopic static dielectric response ($\epsilon_{0}=\epsilon_{\infty}+\epsilon_{ionic}+\epsilon_{other}$). 
A molecular response ($\epsilon_{molecular}$) can occur for materials containing molecules with a permanent dipole, which will likely occur more slowly (due to the moment of inertia of the molecules, and kinetically limited reordering of domains). 
This orientational effect is usually reserved for polar liquids.\cite{van-1937}

We have investigated the energetics of rotation of three organic cations within the lead iodide perovskite structure: (i) ammonium, \ce{NH4+} (A); (ii) methylammonium, \ce{CH3NH3+} (MA); (iii) formamidinium, \ce{NH2CHNH2+} (FA). These were performed using density functional theory (DFT), using the PBEsol\cite{pbesol} exchange-correlation functional and the VASP code\cite{vasp1,vasp2} with the set-up details previously reported.\cite{brivio-042111} Here, we kept the cell lattice parameters fixed and rotated the cell over the long axis of the molecule, an equivalent rotation to tumbling the molecule end-over-end. This gives us an upper, unrelaxed, limit of the rotation barrier. The resulting barriers for rotation in the cage are 0.3, 1.3 and 13.9 kJ/mol, respectively. The value for MA is consistent with observed high rates of rotation at room temperature from $^2$H and $^{14}$N spectra.\cite{wasylishen-58}

The organic cations MA and FA have a large built-in polarisation, most obviously in the case of methylammonium. To investigate this we calculate the polarisation tensor in vacuum with the GAUSSIAN\cite{g09} package on singly charged cations. 
We find that the molecular polarisation tensor is dominated by a dipole contribution. The dipoles, in Debye, for B3LYP/6-31G* (CCSD/cc-pVQZ) calculations are: (i) A, 0.0 (0.0); (ii) MA, 2.29 (2.18); (iii) FA, 0.21 (0.16). An obvious route to increasing the strength of this dipole is successive fluorination of the methyl in methylammonium. We calculate the dipole increase from methylammonium (2.29 Debye) to mono-, bi- and trifluorination to be 5.35, 6.08 and 6.58 Debye, respectively (B3LYP/6-31G*). 
These permanent dipoles will interact with an external electric field, and with each other.

In the hybrid perovskite, the cations are surrounded by a polarisable medium (the perovskite cage), whereas our dipole calculations are in the gas phase. 
As a first approximation we therefore repeat these calculations with the Polarizable Continuum Model (PCM), with a choice of solvent (ethanol, $\epsilon_0=24.852$) which matches our calculated dielectric constant for the bulk material, and is a suitably bulky solvent that it should have a comparable cavity volume to the pore in the perovskite cage.
We use the gas phase geometries. 
The dipoles, in Debye, for B3LYP/6-31G* PCM calculations with ethanol are: (i) A, 0.0; (ii) MA, 2.65; (iii) FA, 0.24; (iv) MA-F3, 7.19. 
A more careful calculation would require a better model for the cavity; however these data do show that only a small deviation from the gas phase values occur, which suggests that the permanent dipole moment of these molecules is robust to the local polarisation environment.

\begin{figure}[ht!]
\begin{center}
\resizebox{7.5 cm}{!}{\includegraphics*{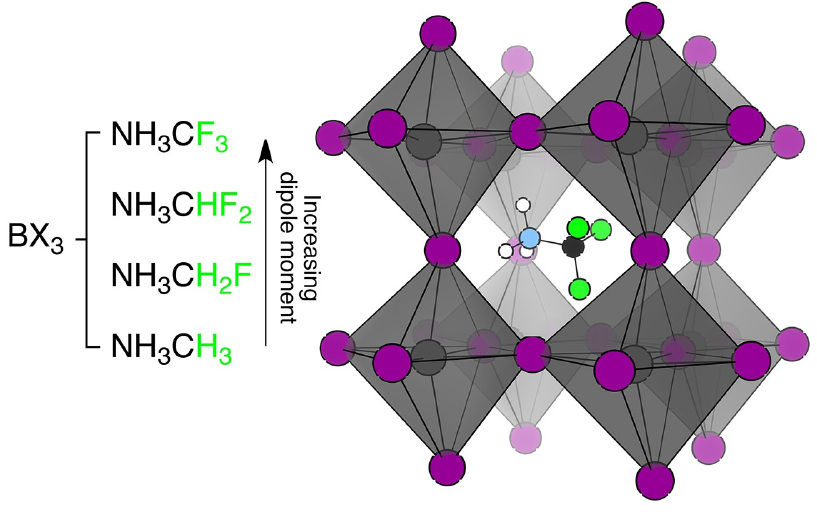}}
\caption{\label{fig-fluor} A significant increase in dipole moment (from 2.3 to 6.6 Debye, by a vacuum calculation with B3LYP/6-31G*) is achieved through increasing the degree of methyl-fluorination. 
} 
\end{center}
\end{figure}

Hybrid halide perovskite solar cells have a typical active region thickness of \SI{1}{\micro\metre}, and a built in voltage of \SI{1}{\volt}, resulting in an electric field of ca. \SI{1e6}{\volt\metre^{-1}}. The dipole potential energy is $\textrm{U}=-\mathbf{p}.\textrm{E}$, and so the energy change in going from aligned to anti-aligned with the electric field is $2.\mathbf{p}.\textrm{E}$ or \SI{4e-5}{\eV D^{-1}}. 
The interaction energy of \SI{9e-2}{\milli\eV} for MA is small compared to the thermal energy, and so weak macroscopic ordering of the cations as a result of the internal field of the solar cell is expected, but more careful quantification is required to understand the field interactions with domains.  

The interaction energy of two point dipoles, going from parallel ferroelectric to anti-ferroelectric alignment, is

\begin{equation}
    U=\frac{2.\mathbf{p}^2}{4\pi\epsilon_0.r^3} 
\end{equation}

In applying this to nearest-neighbour cation dipoles within the perovskite structure we are making two approximations. The first is the point dipole approximation. Our molecules are too close for this to be correct and the origin of the molecular dipole is likely to move within the perovskite cage as the cation rotates. The second is in assuming the vacuum dielectric constant. There will be electron density shielding the region of space between these two cations, but the cations face each other through a relatively open face of the perovskite structure; the upper limit of the screening is the macroscopic static
dielectric constant ($\epsilon_0$ = 25.7 for \ce{MAPbI3})\cite{brivio-042111} that should be valid over longer distances. 

Due to the many unknowns in the location and dynamic behaviour of the cation and the difficulty in fully treating the interaction of the molecular charge density with the full periodic electronic structure of the perovskite, we present these interaction energies as a first approximation. 
These interaction energies can be used to parameterise a classical Hamiltonian for the interaction of the rotating dipoles, and as activation energies for use in looking for order-disorder transitions experimentally. 

Considering a cubic perovskite structure with \SI{6.29}{\angstrom} between nearest neighbours, the parallel ferroelectric alignment energy of methylammonium is \SI{48}{\milli\electronvolt}. Due to the $p^2$ dependence on the dipole moment while keeping methylammonium lattice parameters, this energy is much smaller than thermal energy for FA, \SI{0.3}{\milli\electronvolt}; zero for A; and much greater for fluorinated MA, \SI{436}{\milli\electronvolt}.

Systematic variation of the organic cation (such as by fluorination) and studying the electrical properties of the resulting devices would offer direct evidence of the molecular dipole derived contribution to the operation of hybrid halide perovskite solar cells. Polarisation of the inorganic perovskite cage will be discussed separately below. 

\textbf{Electronic Structure}
For MAPbI$_3$, and related materials, the upper valence band is predominately composed of I 5$p$ orbitals, while the conduction band is formed of Pb 6$p$.\cite{brivio-2014,brivio-042111}
There is hybridisation between the filled Pb 6$s$ band with I 5$p$ that result in anti-bonding states at the top of the valence band (analogous to $p-d$ coupling in II-VI semiconductors). 
Electron and hole conduction therefore occurs along distinct pathways at the \textit{R} point of the first Brillouin zone.\cite{brivio-2014}
Both Pb and I are heavy ions, and as such both the valence and conduction bands contain considerable relativistic effects, i.e. for a quantitative treatment of the electronic band structure spin-orbit coupling must be included.\cite{brivio-2014}
The band gap predicted by a relativistic self-consistent quasiparticle \textit{GW} method is 1.67 eV, in good agreement with the measured value of 1.61 eV from room temperature photoluminescence (PL).\cite{yamada-032302}

Lead is commonly used in ferroelectric and multiferroic materials as its lone pair of electrons is a driving force for structural distortions.\cite{walsh-4455,spaldin-21} Note that the multi-valency of Pb (and Sn for stannous halides) will also be important for the defect chemistry. While present as divalent cations in these structures ($s^2p^0$), a higher tetravalent oxidation state is also accessible ($s^0p^0$). Self-oxidation could facilitate the generation of electrons (n-type doping; $Pb^{II} \rightarrow Pb^{IV} + 2e^-$) or in the trapping of holes (p-type compensation; $Pb^{II} + 2h^+ \rightarrow Pb^{IV}$). The thermodynamics of these processes merit further investigation. 

For all cases investigated, the organic cation acts as a charge compensating centre but does not participate in the frontier electronic band structure.\cite{borriello-235214, brivio-042111} Occupied molecular states are found well below the top of the valence band and empty molecular states found well above the bottom of the conduction band. The choice of organic moiety does influence the lattice constant and, therefore, will change the electronic structure indirectly, in a similar manner to the application of hydrostatic pressure. 

Unusually, the band gaps of hybrid perovskites have been reported to increase with \textit{increasing} cell volume (decreasing pressure),\cite{borriello-235214} which is opposite to the behaviour of most semiconductors.\cite{wei-5404} Indeed the calculated band gap deformation potential,
\begin{equation}
\label{eqn1} \alpha_V = \frac{\partial E_g}{\partial ln V},
\end{equation}
for MAPbI$_3$ is positive ($\alpha_V^R$ = 2.45 eV). As the fundamental band gap is determined at the boundary of the Brillouin zone (\textit{R}), the out-of-phase band-edge states are stabilised as the lattice expands. 
Similar behaviour is observed for Pb chalcogenides.\cite{mark-pbte}
The deformation of the gap at the $\Gamma$ point is negative ($\alpha_V^{\Gamma}$ = -1.08 eV), following standard expectation. 
Measurements of temperature dependent PL indicate a decrease in band gap with decreasing temperature (lattice contraction) to $\sim$ 1.55 eV at 150 K, which is at the onset of a phase change.\cite{yamada-032302} 

The implication of the deformation potential is that for smaller molecular cations, lower band gaps should be observed. Indeed, the smallest possible counter ion is a proton (\ce{H+}), to form \ce{HPbI3}, which results in a calculated lattice parameter  of 6.05 \AA ~ and an associated band gap of less than 0.3 eV. In contrast, the band gap changes on halide substitution are influenced more by the electronic states of the anion, i.e. from Cl to Br to I the valence band composition changes from 3\textit{p} to 4\textit{p} to 5\textit{p} with a monotonic decrease in electron binding energy (lower ionisation potential). The substitution of Br by I in \ce{FAPbX3} has been shown to decrease the band gap from 2.23 eV to 1.48 eV.\cite{snaith-form} 

The effective masses of both carriers are \textit{k}-dependent, as the result of non-parabolic energy bands near the extremal points, as confirmed by self-consistent \textit{GW} calculations.\cite{brivio-2014}
These calculations have also shown that relativistic renormalisation of the band gap is of a similar magnitude to the band gap itself; these materials are far removed from conventional semiconductors such as Si. 
Close to the band extrema the carriers are light with $m_h^*$  and $m_e^*$ of approximately 0.12 $m_0$ and 0.15 $m_0$, respectively.\cite{emass} 

Due to the high dielectric constant, the associated Wannier-Mott exciton radius (within effective mass theory, i.e. $a_0=\frac{\epsilon_0}{m_e}$ in a.u.)\cite{yu-05} is large (204 \AA) with a binding energy ($E_b=\frac{1}{2 \epsilon_0 a_0}$) of 0.7 meV. It is clear that excitons will \textit{not} play a significant role in the photovoltaic device physics. 
Indeed, magnetoabsorption measurements at 4.2 K suggested an exciton radius of 28 \AA ~ and a binding energy of 37 meV, in the low-temperature low-symmetry phase of single-crystal \ce{MAPbI3}.\cite{hirasawa-427}

\begin{figure}[ht!]
\begin{center}
\resizebox{8 cm}{!}{\includegraphics*{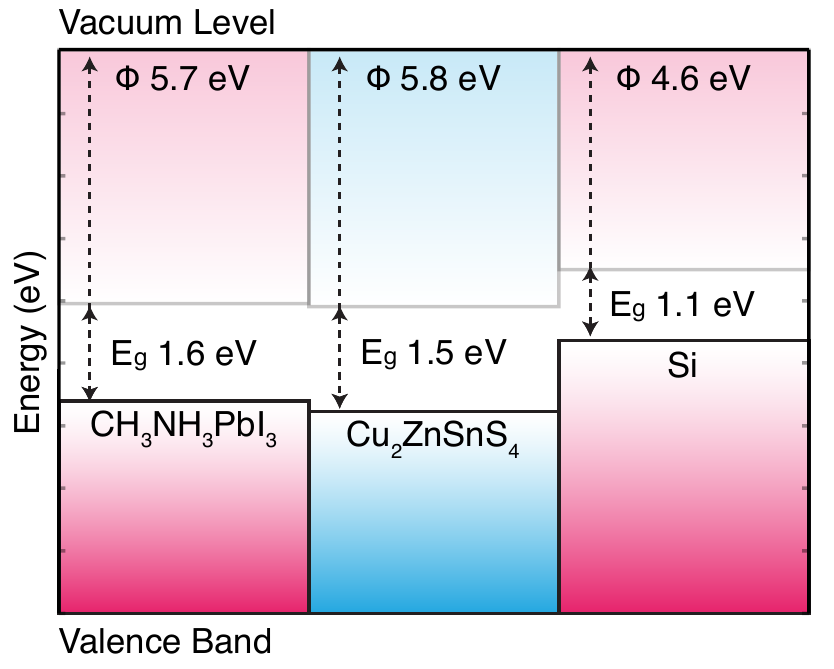}}
\caption{\label{fig-offset} Valence band ionisation potentials of \ce{MAPbI3} with respect to the vacuum level. Calculations were performed on the non-polar (110) surface with slab thickness of 25 \AA ~ and a vacuum thickness of 15 \AA. The Kohn-Sham eigenvalues (PBEsol) are corrected by the bulk quasi-particle energies ($\Delta$E=0.2 eV). Values for the energetically similar inorganic thin-film absorber \ce{Cu2ZnSnS4}\cite{walsh-400}, and Si\cite{crc} are shown for comparison. 
} 
\end{center}
\end{figure}

The valence band energy (ionisation potential) of \ce{MAPbI3} is 5.7 eV below the vacuum level (Figure \ref{fig-offset}). 
The computed bulk ionisation potential and electron affinity of 5.7 eV and 4.1 eV are consistent with the use of TiO$_2$ (electron acceptor) and Au (gold acceptor) in photovoltaic devices.\cite{park-2423}
This ionisation potential is comparable to thin-film solar cell absorbers such as \ce{Cu2ZnSnS4},\cite{todorov-156} which implies that other device configurations, e.g. a Mo back-contact (to replace Au) and a CdS buffer layer (to replace TiO$_2$), are possible. 
In contrast, the valence band of Si is significantly higher in energy. 

\textbf{Carriers, Defects and Decomposition}
The bulk electronic structure also has implications for carrier concentrations. These materials are intrinsic semiconductors with initial reports of relatively high doping densities, e.g. above 10$^{19}$cm$^{-3}$ in \ce{MASnI3}.\cite{mitzi-159}
Unlike traditional inorganic absorber materials where defect formation can be controlled thermodynamically, the solution processing of these materials at low temperatures is more likely to be kinetically controlled with defects in the lattice (e.g. cages with missing MA cations) being a product of the crystallisation process. 
Initial computations suggest that the low energy defects all exhibit resonant or shallow impurity bands.\cite{yin-063903} 

The Mott criterion ($n_c a_0 = 0.26$)\cite{edwards-2575} based on the calculated band structure parameters, predicts a transition to a degenerate semiconductor could happen from carrier concentrations ($n_c$) as low as 10$^{16}$cm$^{-3}$. 
Note that this is a lower estimate which will be increased by both the non-parabolic nature of the band edges and fluctuations in the local elecrostatic potentials due to structural inhomogeneity. 

There are a number of potential routes for the control of carrier concentrations (doping). Firstly, a mixture of charged and neutral counter-cations (e.g. \ce{NH4+} and \ce{NH3}) would result in an electron deficiency and hence p-type doping. Conversely, the inorganic change could be altered by partial substitution of Pb by a trivalent cation would result in an electron excess and hence n-type doping. Tl and Bi are two obvious candidates due to their similar size.

\begin{figure}[!]
\begin{center}
\resizebox{8.3 cm}{!}{\includegraphics*{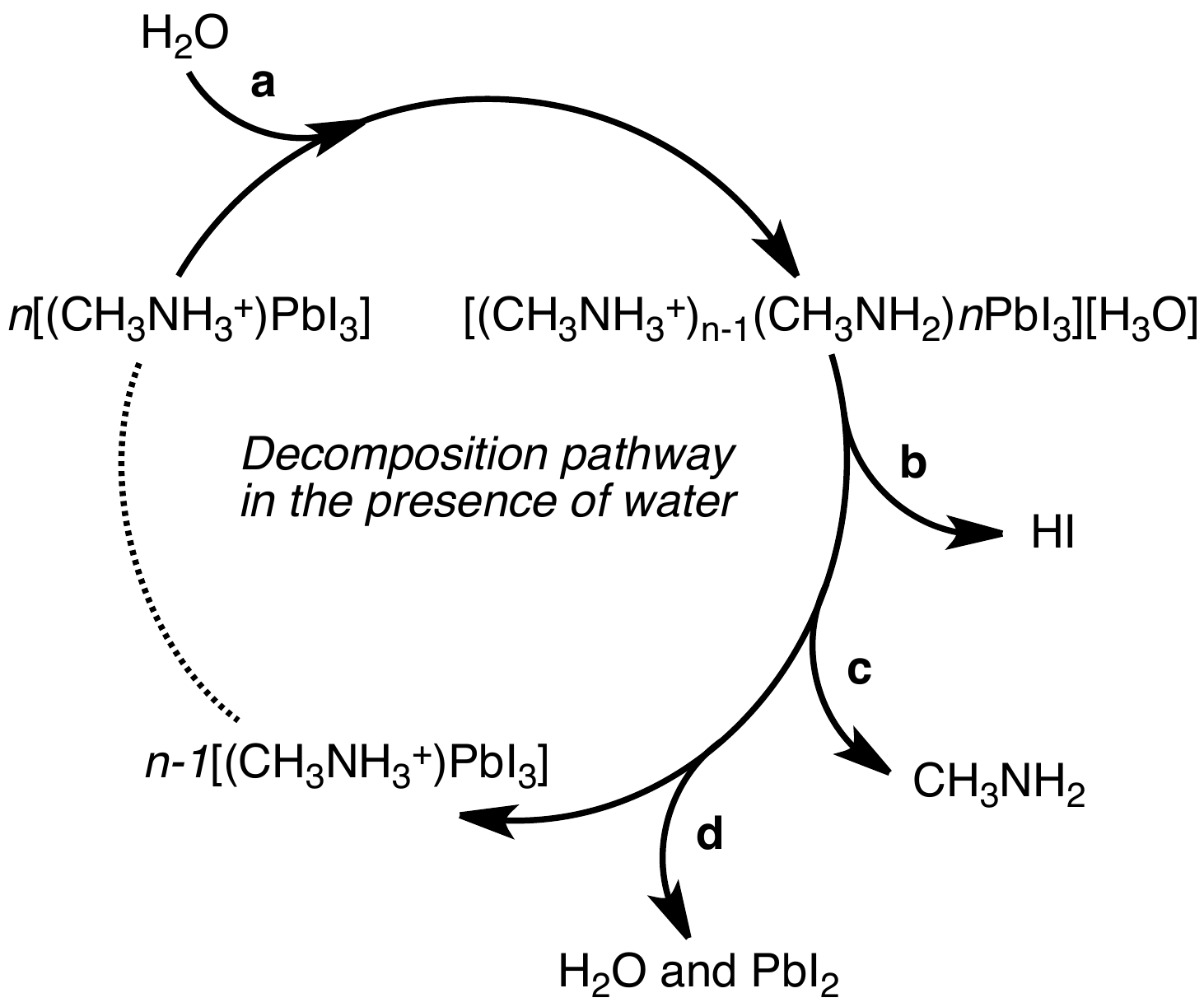}}
\caption{\label{fig-decomp} Possible decomposition pathway of hybrid halide perovskites in the presence of water.  One water molecule is required to initiate this process, with the decomposition being driven by the phase changes of both hydrogen iodide (soluble in water) and the methylammonia (both volatile and soluble in water).  
} 
\end{center}
\end{figure}

There are reports of hybrid perovskites reacting with Lewis bases; the most notable being irreversible degradation in the presence of \ce{H2O} and temporary bleaching in the presence of ammonia.\cite{zhao_optical_2014} 
There are several plausible mechanisms by which this decomposition may occur. We propose the simple acid-base reaction shown in Figure \ref{fig-decomp}. 
In the case of water exposure, a single water molecule is sufficient to degrade the material; however, an excess of water is required to dissolve the HI and \ce{CH3NH2} by-products.  
As a result, a closed system with trace amounts of \ce{H2O} will result in partial decomposition of the hybrid perovskite until either: (i) the HI has saturated the \ce{H2O} or (ii) the vapour pressure of \ce{CH3NH2} has reached equilibrium.  In the presence of sufficient water the material degrades entirely to form \ce{PbI2}.

\begin{figure*}[ht!]
\begin{center}
\resizebox{15 cm}{!}{\includegraphics*{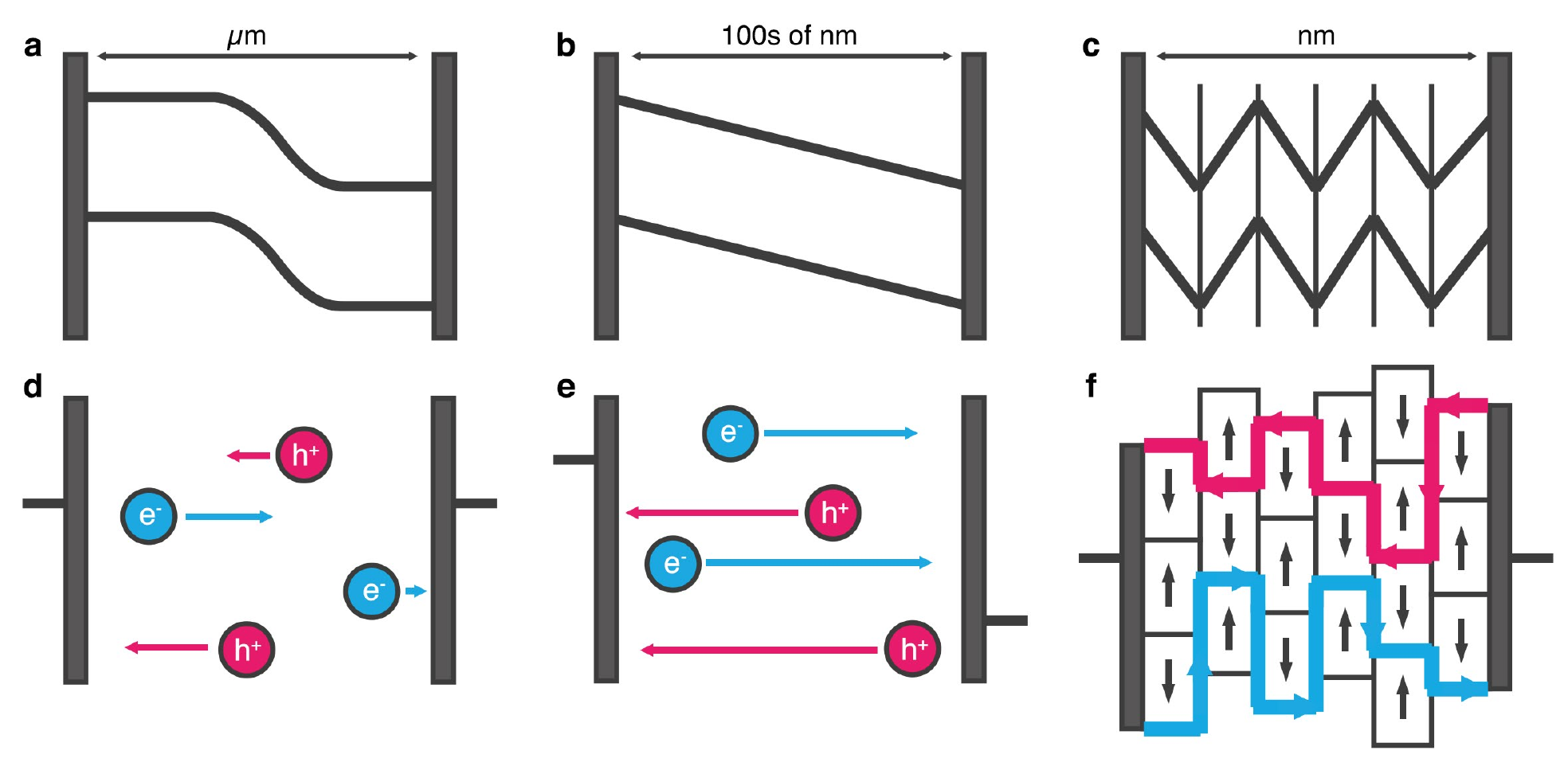}}
\caption{\label{fig-highway} Schematic of the 1D built-in electric potential (upper panels) and associated 2D electron and hole separation pathways (lower panels) in (a,d) a macroscopic p-n junction; (b,e) a single domain ferroelectric thin-film; (c,f) a multi-domain ferroelectric thin-film. In the multi-domain ferroelectric -- which we propose for hybrid perovskites -- electrons will move along minima in the electric potential, while holes will move along maxima (i.e. anti-phase boundaries). 
} 
\end{center}
\end{figure*}

The proposed mechanism for reversible perovskite bleaching is similar to that of water decomposition, where a reaction analogous to Figure \ref{fig-decomp} may occur. In this case, the ionic salt (and strong acid) HI, would form a separate ionic salt with \ce{NH3}.  This Grotthuss mechanism\cite{agmon_grotthuss_1995} should cause a concerted proton transfer throughout the material resulting in three thermally reversible outcomes: (i) the methylammonia evaporates, resulting in the inclusion of \ce{NH4} as the organic ion in the perovskite, with the simultaneous production of HI; (ii) the system degrades to form HI, \ce{PbI2} and volatile organics; (iii) the \ce{CH3NH3+}/\ce{CH3NH2}--\ce{NH3}/\ce{NH4+} reaction occurs on the surface causing the material to form the neutral \ce{CH3NH2}. This acid-base reaction is reversible and ammonia is more volatile than the heavier methylammonia (which may be trapped within the material), resulting in transient bleaching.  

To our knowledge, there has been no reports of hybrid perovskites incorporating aprotic organic ions (such as tetramethylammonium, \ce{(CH3)4N+}). Such a material would not be capable of the reaction mechanism depicted in Figure \ref{fig-decomp}, and so may be more chemically stable. The built-in molecular dipole, which could be important for the performance of these materials in a solar cell, could be achieved by selective flourination.   

We speculate that decomposition or reversible bleaching should be observed in the presence of any \textit{small} Lewis acid with a protic hybrid perovskite, and such an experimental study would help understanding the degradation pathways of hybrid perovskite solar cells.

\begin{table}[hb]
\small
  \caption{Calculated properties of four hybrid lead halide perovskites from density functional theory. 
  The molecular dipole (D) is given in Debye, calculated by vacuum B3LYP/6-31G* in GAUSSIAN. The pseudo-cubic lattice constant ($a = \sqrt[3]{V}$) in \AA;  the lattice electronic  polarisation ($\Delta P$) in $\mu$C/cm$^2$; and the rotation barrier ($E_{rot}$), calculated by PBESol in VASP. The nearest-neighbour dipole interaction ($E_{dip}$) in kJ/mol is estimated from a point dipole calculation.}
  \label{tbl:properties2}
  \begin{tabular}{ld{2}d{2}d{0}d{1}d{2}}
    \hline
    Cation        &  D    &  a  & \Delta P  & E_{rot} & E_{dip}\\
    \hline
    \ce{NH4}      &  0.00 & 6.21  &  8  & 0.3   & 0.00 \\
    \ce{CH3NH3}   &  2.29 & 6.29  &  38 & 1.3   & 4.60 \\
    \ce{CF3NH3}   &  6.58 & 6.35  &  48 & -     & 42.00 \\
    \ce{NH2CHNH2} &  0.21 & 6.34  &  63 & 13.9  & 0.03 \\ 
    \hline
\end{tabular}
\end{table}

\textbf{Spontaneous Electric Polarisation}
Most inorganic perovskites display spontaneous electric polarization, arising from the breaking of crystal centro-symmetry, as a result of the \textit{B} cation moving away from the centre of the $BX_6$ octahedron.\cite{glazer-1972} The phenomenon is particularly pronounced in hybrid halide perovskites, where the asymmetry of the organic cation ensures the absence of an inversion centre in the structure. 

The magnitude of the bulk polarisation has been probed using Berry phase calculations within the modern theory of polarisation.\cite{berry1,berry2} The calculated values (PBEsol functional) are presented in Table \ref{tbl:properties2}.
For \ce{MAPbI3}, the electronic polarisation of 38 $\mu$C/cm$^2$ is comparable to ferroelectric oxide perovskites (e.g. ca. 30  $\mu$C/cm$^2$ in \ce{KNbO3}\cite{dall-10105}).
Indeed, the application of ferroelectric oxides to photovoltaics has been recently demonstrated.\cite{grinberg-509}

The strong polarisation of the lattice has two potential advantages for photovoltaic operation: (i) enhanced charge separation and concomitant improved carrier lifetimes;\cite{yang-143} (ii) open circuit voltages above the band gap of the material.\cite{grinberg-509} Both effects are linked to the internal electric field, which results as a consequence of lattice polarization. The mechanisms are shown schematically in Figure \ref{fig-highway}.

Traditional semiconductor photovoltaic devices separate charge carriers at p-n junctions (Figure \ref{fig-highway} a). A ferroelectric domain, with its built-in electric field (Figure \ref{fig-highway} b), acts to separate the exciton generated by photo-absorption, in effect acting as a p-n junction. 
A standard planar p-n junction is on the order of micrometers and separated charge carriers must diffuse through the junction to reach their respective electrodes; during this diffusion process carriers may encounter one another and recombine. 
The size of ferroelectric domains is smaller, on the nanometer scale (several cages for the hybrid perovskites). 
The probability of carrier recombination within a domain is therefore reduced with respect to a traditional heterojunction. 

The carriers in the ferroelectric material can diffuse to domain boundaries, where there is a build up of carriers of a given type (Figure \ref{fig-highway} c) along peaks and troughs in the electric potential generated by local dipole order. 
The carriers may then diffuse along these `ferroelectric highways' toward the electrodes, unimpeded by carriers of the opposite charge; essentially we have an intrinsic semiconductor region where local hole and electron segregation reduces recombination. 
We propose this as the origin of exceptionally long carrier diffusion lengths\cite{snaith-6156,gratzel-6156} despite local structural disorder.
The carrier density of these ferroelectric boundaries as well as the domain orientation will be influenced by the applied voltage and hence give rise to hysteresis, or spurious fluctations, in electrical measurements.

The contribution of these photoferroic effects could be assessed and controlled by modification of the counter ion. 
The molecular dipole can be tuned, with the simple example discussed earlier being the incorporation of more electronegative species on one side of the MA cation (Figure \ref{fig-fluor}).
Even for \ce{APbI3}, where there is no molecular dipole the lattice is weakly polarised (8 $\mu$C/cm$^2$) and this effect would likely be enhanced towards lower symmetry interfaces and grain boundaries. Notably, the largest degree of polarisation is observed for  \ce{FAPbI3} (63 $\mu$C/cm$^2$), which is not the most polar cation, but its larger size induces a polar deformation of the \ce{PbI3} cage.

It is interesting to note that the `ferroelectric highways' are distinct but similar to the proposed mechanism for carrier separation in thin-film \ce{CuInSe2} solar cells, where Cu-rich and Cu-poor domains (ordered defect complexes) have been linked to internal homojunctions that facilitate efficient charge carrier separation and reduced carrier recombination.\cite{zhang-9642,persson_anomalous_2003}

~
In summary, we have provided insights into the key materials properties of organometallic perovskites that contribute to their photovoltaic performance. Further investigations are required to assess the  mechanisms proposed including the role of ferroelectric domains and the small exciton binding energies. Opportunities have been discussed for rational modification of material characteristics through modifying the organic cation. In particular we suggest manipulating the dipole moment (to affect ferroelectric behaviour and dielectric constant), cation size (to manipulate band gap, and ferroelectric behaviour) and to move to an aprotic cation (to avoid a degradation pathway). 
A lead free material with high performance and reduced degradation would represent a major step forward for this technology. We hope that the description of the atomistic origin of favourable material characteristics exhibited by \ce{MAPI3} may offer some guidance for the design of improved organometallic perovskites.

\begin{acknowledgement}

We thank L.M. Peter and P.J. Cameron for useful discussions, and acknowledge membership of the UK's HPC Materials Chemistry Consortium, which is funded by EPSRC grant EP/F067496. 
J.M.F. and K.T.B. are funded by EPSRC Grants EP/K016288/1 and EP/J017361/1, respectively.
F.B. is funded through the EU DESTINY Network (Grant 316494).
C.H.H. is funded by ERC (Grant 277757). A.W. acknowledges support from the Royal Society. 
The authors declare no competiting financial interests. 

\end{acknowledgement}

\begin{suppinfo}
    GAUSSIAN input and output files used in this work are electronically available in a Figshare dataset.\cite{figshare}
\end{suppinfo}

\bibliography{library}

\end{document}